\documentclass[hyper,letterpaper,notoc]{JHEP}
\newcommand{\on}[2]{\,(\!\!\begin{array}{c}  
                        {#1}                
                        \\                  
                        {#2}                
                        \end{array}\!\!)} 

\newcommand{\D}{{\cal D}}
\newcommand{\G}{{\cal G}}
\newcommand{\N}{{\cal N}}

\newcommand{\R}{{\cal R}}
\newcommand{\Z}{{Z}}
\preprint{TAUP-2656-2000 \\ {\tt hep-th/0101017}}

\title{Superconformal Mechanics and the Super 
Virasoro Algebra}

\author{Alon Marcus
\\
Raymond and Beverly Sackler Faculty of Exact Sciences\\
 School of Physics and Astronomy\\
 Tel Aviv University, Ramat Aviv, 69978, Israel\\
\email{alon@post.tau.ac.il}
}

\abstract{ We consider ${\cal N}=1,2$ superconformal mechanics 
in $0+1$ dimensions and show that if the Hamiltonian is 
invertible the superconformal generators can be used to 
construct half of the super Virasoro algebra. 
The full algebra can be derived if the special conformal 
generator is also invertible. The generators are quantized and
a general prescription is given for the construction of the
$\N=1$ algebra independently of the specific details of 
the superconformal mechanics provided that in addition
its quadratic Casimir operator vanishes. }
\keywords{Super conformal mechanics, super Virasoro algebra} 

\begin{document}
\section{Introduction}
There is an ongoing interest in conformal mechanics, since the early work 
of~\cite{Fubini:1976}, and in the supersymmetric 
generalization~\cite{Fubini:1984,Akulov:1984uh,Deotto:2000hz}.
These one dimensional conformal field theories admit exact 
solutions to problems that can be accessed only perturbatively 
in higher dimensions, due to the existence
of the powerful conformal symmetry that constrains the dynamics. 
Although these $d=1$ conformal mechanics are relatively simple 
they are still not trivial.
A geometrical picture that relates the one dimensional field equations
to geodesics in the group space of $SO(1,2)$ and $SU(1,1|1)/U(1)$
was constructed in~\cite{Ivanov:1989vw,Ivanov:1989it}.

Recently, the $AdS_{p+2}/CFT_{p+1}$ conjecture, 
see e.g~\cite{Aharony:1999ti,Cadoni:1999ja}, has
raised a new interest in conformal mechanics, i.e the case of $p=0$.
This conjecture proposes that in an appropriate limit certain conformal
field theories in $p+1$ dimensions are dual to superstring theory on 
an $AdS$ space in $p+2$ dimensions times some compact manifold. 
The case of conformal quantum mechanics and the corresponding 
$AdS_2$~\cite{Strominger:1998yg} may offer more insights to this 
conjecture due to the simplification of an
analysis in so few dimensions, though there are also new features since
the boundary of the $AdS_2$ is not connected. 

Although the basis to this paper is mainly algebraic,
it is instructive to bear in mind concrete realizations of 
physical systems which are governed by (super) conformal 
mechanics. An illustrative example 
is that of the physics of a test particle in the near horizon region 
of a $d=4$ $\N=2$ extreme Reissner-Nordstr\"om black 
hole~\cite{Claus:1998ts,deAzcarraga:1998ni}.
This particle is described by conformal mechanics where the 
canonical coordinates are the radial coordinate and radial 
momentum in $d=4$. The main interest in such systems
is due to the fact that black holes provide the arena where 
gravity and quantum mechanics match~\cite{Britto-Pacumio:1999ax}. 
Understanding these systems shed light on the 
intriguing problems of quantum gravity. Another example is that 
of a non-relativistic spinning particle coupled to a magnetic 
field and a scalar potential~\cite{Papadopoulos:2000ka}~\footnote{For 
this system to be (super) conformal invariant one need to make 
diffeomorphisms transformations on the background to compensate on the 
conformal transformations.}.

In this paper we follow the analysis in~\cite{Kumar:1999fx,Kumar:2000at} 
on conformal symmetry and the Virasoro algebra and extend it to the case
of $\N=1,2$ super conformal symmetry. In section~\ref{sqm_gen} we
identify the superconformal algebra as the subalgebra of the
super Virasoro algebra. In 
section~\ref{clas_alg} we obtain classical recursion equations
for the generators 
and find a representation of the full algebra for a free particle and
an interacting one. At the level of Poisson brackets this algebra close
to $\N=1,2$ Neveu-Schwarz super Virasoro 
algebra, with a $U(1)$ Kac-Moody algebra~\cite{Ragoucy:1992aq} 
for the latter case. In section~\ref{quantization} we describe the 
quantization of the generators and give a general construction 
of $\N=1$ super Virasoro generators out of the superconformal ones.

\section{Generators of Superconformal Quantum Mechanics}
\label{sqm_gen}
The generators of $0+1$ dimensional superconformal quantum mechanics obey 
the super algebra of $osp(2|2)\cong 
su(1,1|1)$~\cite{deAzcarraga:1998ni}. Their non-trivial commutation and 
anti-commutation relations can be written in the following 
way~\footnote{We rescaled the fermionic charges by a factor $\sqrt{2}$
relative to~\cite{deAzcarraga:1998ni}.}
\begin{equation}
\label{scalgebra}
\begin{array}{lll}
[H,D]_-=iH \quad, & [K,D]_-=-iK \quad, & [H,K]_-=2iD \quad,
\\[0.3cm]
[Q_i,Q_j]_+=2\delta_{ij} H \quad,
& [S_i,S_j]_+=2\delta_{ij} K \quad, &
\displaystyle [Q_i,S_j]_+=2\delta_{ij} D +
\epsilon_{ij} B \quad,
\\[0.3cm]
\displaystyle [D,Q_i]_-= -\frac i2 Q_i \quad,
& \displaystyle [D,S_i]_-= \frac i2 S_i \quad,
\\[0.3cm]
[K,Q_i]_-= -i S_i \quad, & [H,S_i]_-= i Q_i \quad,
\\[0.3cm]
[B,Q_i]_-= -i \epsilon_{ij} Q_j \quad,
& [B,S_i]_-= -i \epsilon_{ij} S_j \quad, & i,j=1,2 \quad ,
\end{array}
\label{1.4}
\end{equation}
where $[\,,\,]_\pm$ stands for commutators and anticommutators.
The following identification of the generators
\begin{eqnarray}
\begin{array}{lll}
L_{-1}=H \quad, & L_0=-D \quad, & L_1=K \quad,
\nonumber
\\
\\
\nonumber
G^i_{-1/2}=Q_i   \quad,  & G^i_{1/2}=-S_i  \quad , & B_0=B 
\end{array}
\end{eqnarray}
can be used to recast the algebra~(\ref{scalgebra}) as
a subalgebra of the super Virasoro algebra with a $U(1)$ charge
\begin{eqnarray}
\label{quantum_algebra}
\nonumber
[L_m,L_n]_- & = & i(m-n)L_{m+n}
\\
\nonumber
\,[L_m,G^i_r]_- & = & i(\frac{1}{2}m-r)G^i_{m+r}
\\
\,[G^i_r,G^j_s]_+ & = & 2\delta_{ij}L_{r+s}+
\epsilon_{ij}(r-s)B_{r+s}
\\
\nonumber
\,[B_k,L_m]_- & = & i k B_{k+m} 
\\
\nonumber
\,[B_k,G^i_r]_- & = & -i\epsilon^{ij}G^j_{k+r}
\end{eqnarray}
for $m,n=-1,0,1$, $r,s=-\frac{1}{2},\frac{1}{2}$ and $k=0$.
A realization of this algebra is given for the theory of a free particle 
by
\begin{eqnarray}
\label{free_particle}
\begin{array}{lll}
K=\frac{1}{2}x^2 \quad,  & D=-\frac{1}{4}(xp+px) \quad,  & H=\frac{1}{2}p^2
\\
Q^i=\psi^i p \quad, & 
S^i=-\psi^i x \quad, & 
B=\frac{i}{2}[\psi^1,\psi^2]_-
\end{array}
\end{eqnarray}
where $\psi^i$, $i=1,2$ are Grassmann coordinates 
$[\psi^i,\psi^j]_+=\delta^{ij}$. 

For completion we give an explicit form of the conserved charges
associated with this algebra written in a way that is compatible 
with the whole super Virasoro algebra. Time translation is generated by
$H=L_{-1}$ so for any quantum generators $\G$ the Heisenberg 
representation gives 
\begin{eqnarray}
\label{heisenberg}
\G(t)=e^{iL_{-1}t}\G e^{-iL_{-1}t} 
= \sum_{l=0}^\infty \frac{(-it)^l}{l!}
[[\G,L_{-1}],L_{-1}],...,L_{-1}] \, ,
\end{eqnarray}
where the first term corresponding to $l=0$ is 
$\G$. 
The nested commutators are easily calculated for any super Virasoro generator
\begin{eqnarray}
\label{conserved_charges}
\nonumber
G_r^i(t) & = & \sum_{l=0}^\infty  \on{r+1/2}{l} G^i_{r-l}t^l 
\\
L_n(t)   & = & \sum_{l=0}^\infty  \on{n+1}{l} L_{n-l}t^l 
\\
\nonumber
B_k (t)  & = & \sum_{l=0}^\infty  \on{k}{l} B_{k-l}t^l \, ,
\end{eqnarray}
which are finite sums for $r\ge -1/2$, $n\ge -1$ and $k\ge 0$.
If, for example, $k=-m<0$ we may use analytic continuation
\begin{eqnarray}
\nonumber
l!\on{k}{l}\rightarrow \lim_{\epsilon\rightarrow 0}
\frac{\Gamma(-m+1+\epsilon)}{\Gamma(-m-l+1+\epsilon)}=
(-1)^l\frac{\Gamma(m+l)}{\Gamma(m)} \, .
\end{eqnarray}
The generators in~(\ref{conserved_charges}) are conserved by construction
\begin{eqnarray}
\frac{d\G}{d\,t}=i[\G,L_{-1}]+
\frac{\partial\G}{\partial\,t}=0 \,.
\end{eqnarray}
\section{Classical algebra}
\label{clas_alg}
In order to find the complete super Virasoro algebra we use 
the classical algebra
\begin{eqnarray}
\label{classical_algebra}
\nonumber
\{L_m,L_n\} & = & (m-n)L_{m+n}
\\
\nonumber
\{L_m,G^i_r\} & = & (\frac{1}{2}m-r)G^i_{m+r}
\\
\{G^i_r,G^j_s\} & = & 2\delta_{ij}L_{r+s}+
\epsilon_{ij}(r-s)B_{r+s}
\\
\nonumber
\{B_k,L_m\} & = & k B_{k+m} 
\\
\nonumber
\{B_k,G^i_r\} & = & -\epsilon^{ij}G^j_{k+r}
\\
\nonumber
\{B_k,B_m\} & = & 0   \, ,
\end{eqnarray}
where $\{,\}$ denote the even Poisson 
Brackets~\cite{Khudaverdian:1991xa}
\begin{eqnarray}
\{G,K\}=\frac{\partial G}{\partial x}\frac{\partial K}{\partial p}
-\frac{\partial G}{\partial p}\frac{\partial K}{\partial x}-
\sum_{l=1}^2(-1)^{p(G)}\frac{\partial G}{\partial \theta^l}
\frac{\partial K}{\partial \theta^l} \quad ,
\end{eqnarray}
with $p(Q)=1$ for an odd generator and zero otherwise.
This definition gives for even coordinates $xp-px=0$
\begin{eqnarray}
\{x,p\}=1
\end{eqnarray}
and for odd coordinates $\theta^i\theta^j+\theta^j\theta^i=0$
\begin{eqnarray}
\{\theta^i,\theta^j\}=\delta^{ij} \, .
\end{eqnarray}

In what will follow we use the explicit 
representation of the global 
generators, e.g the generators that are given in 
equation~(\ref{free_particle}), and the classical algebra 
to obtain a set of solvable differential equations. 
This is done by calculating the Poisson Brackets of a general generator 
of the algebra with one of the global generators and demanding that it will 
be equal to the r.h.s of the algebra~(\ref{classical_algebra}), 
as was done in~\cite{Kumar:1999fx} for
the $L_n$ generators.
\subsection{Free Particle}
We begin with the super conformal generators of the free 
particle that are given in equation~(\ref{free_particle}).
The Poisson Brackets of the $G_r^i$ generators with the Hamiltonian 
and the special conformal generator gives
\begin{eqnarray}
-p\frac{\partial G_r^i}{\partial x} & = & \{H,G_r^i\}=
   \{L_{-1},G_r^i\} = (-\frac{1}{2}-r)G_{r-1}^i
\nonumber
\\
\\
\nonumber
x\frac{\partial G_r^i}{\partial p} & = & \{K,G_r^i\}=
   \{L_1,G_r^i\} = (\frac{1}{2}-r)G_{r+1}^i             \quad .
\end{eqnarray}
The integrability condition
$\frac{\partial^2 G_r^i}{\partial p\partial x}=
\frac{\partial^2 G_r^i}{\partial x\partial p}$    
gives the recursion formula for $ G_r^i$
\begin{eqnarray}
\frac{1}{x^2}(r-\frac{1}{2})G^i_{r+1}=-\frac{1}{p^2}(r+\frac{1}{2})G^i_{r-1}
+\frac{2r}{xp}G_r^i         \quad .
\end{eqnarray}
Similar equations for $L_n$ and $B_k$ exist and are easily solved. 
The classical Virasoro generators are then given by
\begin{eqnarray}
\label{free_gen}
\nonumber
G_r^i & = & \theta^i x^{\frac{1}{2}+r}
p^{\frac{1}{2}-r}
\\
L_n   & = & \frac{1}{2}x^{1+n}p^{1-n}
\\
\nonumber
B_k   & = & B\, x^k p^{-k} 
\end{eqnarray}
for $n,k\in \Z$ and $r\in \Z+\frac{1}{2}$, 
$B=\frac{1}{2}[\theta^1,\theta^2]$.
For $n\ge -1$, $r\ge -1/2$ and $k\ge 0$ 
we can recast these generators as
\begin{eqnarray}
\label{multiplied_generators}
L_n   & = & L_{-1}(L_0L_{-1}^{-1})^{n+1}
\nonumber
\\
G_r^i & = & G^i_{-1/2}(L_0L_{-1}^{-1})^{r+1/2}
\\
\nonumber
B_k   & = & B_0(L_0L_{-1}^{-1})^k \quad .
\end{eqnarray}
This is an extension to~\cite{Kumar:2000at}. The representation, however, 
is not ``unitary'' $L_n^*\ne L_{-n}$. To get a unitary representation
we consider the linear combination of DFF~\cite{Fubini:1976}
\begin{eqnarray}
\nonumber
L_1 & = & \frac{1}{2}(a H-\frac{K}{a}- 2iD)  =  \frac{1}{2}z^2
\\
L_0 & = & \frac{1}{2}(a H+\frac{K}{a})  =  \frac{1}{2}z z^*
\\
\nonumber
L_{-1} & = & \frac{1}{2}(a H-\frac{K}{a}+ 2iD)  =  \frac{1}{2}{z^*}^2 \, ,
\end{eqnarray}
where we define 
\begin{eqnarray}
\label{coordinate_transformation}
z=\frac{\sqrt{a}p-i \frac{x}{\sqrt{a}}}{\sqrt{2}}
&  \,  &
z^*=\frac{\sqrt{a}p+i \frac{x}{\sqrt{a}}}{\sqrt{2}} \quad .
\end{eqnarray}
To obtain the entire algebra we can solve the differential equations 
again or observe that the new generators have the same form as 
in equation~(\ref{free_gen}) with 
$x\rightarrow z$ and $p\rightarrow z^*$. Therefore the solution
\begin{eqnarray}
\label{unitary_rep}
\nonumber
L_n   & = & \frac{1}{2}z^{1+n}{z^*}^{1-n}
\\
B_k   & = & Bz^k{z^*}^{-k}
\\
\nonumber
G^i_r & = & \theta^i z^{1/2+r}{z^*}^{1/2-r}
\end{eqnarray}
obey the algebra up to new factors of $i$
due to the fact that the coordinate
transformation~(\ref{coordinate_transformation}) is canonical up to a 
scale, i.e $\{z,z^*\}=-i$.
Another difference between the two representations, besides unitarity, is
that for the generators in~(\ref{unitary_rep}) time translation is 
generated by $L_0$ and not by $L_{-1}$.
\subsection{An Interacting Particle}
\label{inter_cl}
To obtain the generators of an interacting particle we note that the
classical Virasoro algebra is invariant under $L_n\rightarrow f^n L_n$ ,
$G^i_r\rightarrow f^r G^i_r$ and $B_k\rightarrow f^k B_k$ where
$f=f(L_0)$. Since for~(\ref{free_gen}) $L_0=2xp$ we get
\begin{eqnarray}
\label{int_gen}
\nonumber
G_r^i & = & \theta^i x^{\frac{1}{2}+r}
p^{\frac{1}{2}-r}f^r
\\
L_n   & = & \frac{1}{2}x^{1+n}p^{1-n}f^n
\\
\nonumber
B_k   & = & B\, x^k p^{-k} f^k \quad .
\end{eqnarray}
The Hamiltonian of a particle in a conformal potential is obtained 
for $f^{-1}=1+\frac{g}{x^2 p^2}$
\begin{eqnarray}
L_{-1}=H=\frac{1}{2}(p^2+\frac{g}{x^2}) \, .
\end{eqnarray}
For a superconformal potential we
first recall the general method~\cite{Witten:1981nf} 
for supersymmetrizing a classical and quantum mechanical Hamiltonian.
Given a Hamiltonian $H=\frac{1}{2}p^2+V$ we can define the odd generators
\begin{eqnarray}
Q^i=(p{\bf 1}+W_{,x}\,{\bf \varepsilon})^{ij}\psi^j \, ,
\end{eqnarray}
where $W_{,x}=\frac{d W}{d x}$, which obey the following relation
\begin{eqnarray}
\{Q^i,Q^j\}=\delta^{ij}\left(p^2+W_{,x}^2-2B\, W_{,xx} 
\right)\, .
\end{eqnarray}
By solving $\sqrt{2V}=W_{,x}$ we get 
\begin{eqnarray}
\{Q^i,Q^j\} & = & 2\delta^{ij}H_{susy}
\\
H_{susy}    & = & \frac{1}{2}\left(p^2+{W_{,x}}^2-2B\, W_{,xx}
\right) \, .
\end{eqnarray} 
The potential for conformal mechanics
is $V=\frac{g^2}{2x^2}$, which is solved for
$W=\frac{1}{2}g\log{x^2}$. 
These expressions for the Hamiltonian and the supersymmetric generators
motivate the following ansatz
\begin{eqnarray}
\nonumber
G_r^i & = &  x^{\frac{1}{2}+r}
p^{\frac{1}{2}-r}f^r(1+\frac{g^2}{u^2})^{-r-1/2}
({\bf 1}+\frac{g}{u}{\bf \varepsilon})^{ij}\psi^j
\\
L_n   & = & \frac{1}{2}x^{1+n}p^{1-n}f^n
(1+\frac{g^2+2gB}{u^2})^{-n}
\\
\nonumber
B_k   & = & x^kp^{-k}f^k(1+\frac{g^2}{u^2})^{-k} B \, ,
\end{eqnarray}
which obey~(\ref{classical_algebra}) and can be recast in the form 
of~(\ref{multiplied_generators}).

One can use the classical analog of~(\ref{heisenberg}) to obtain an explicit 
representation of the conserved charges associated with 
the classical generators, e.g the use of the generators in
equation~(\ref{int_gen}) will give
\begin{eqnarray}
\nonumber
G_r^i(t) & = & \theta^i p/\sqrt{f}(x/p\,f+t)^{r+1/2}
\\
L_n(t)   & = & \frac{1}{2f}p^2(x/p\,f+t)^{n+1}
\\
\nonumber
B_k (t)  & = & B(x/p\,f+t)^k \, ,
\end{eqnarray}
which is the subalgebra of $w_\infty$ obtained in~\cite{Cacciatori:1999rp}.
\section{Quantization of the generators}
\label{quantization}
For the quantization of the generators we first rescale the algebra
and absorb all factors of $i$. For clearness we give the  
$\N=0,1,2$ algebras:
\begin{eqnarray}
\N=0:   &  [L_m,L_n]_-=(m-n)L_{m+n}
\label{N=0}
\\
\nonumber
\\
\nonumber
\N=1:   &  [L_m,L_n]_-=(m-n)L_{m+n}
\\      &  [L_m,G_r]_-=(\frac{1}{2}m-r)G_{m+r}
\label{N=1}
\\      &  [G_r,G_s]_+=2L_{r+s}
\nonumber
\\
\nonumber
\\
\nonumber
\N=2:   &  [L_m,L_n]_-=(m-n)L_{m+n}
\nonumber
\\      &  [L_m,G^i_r]_-=(\frac{1}{2}m-r)G^i_{m+r}
\nonumber
\\      &  [G^i_r,G^j_s]_+=2\delta^{ij}L_{r+s}+(r-s)\epsilon^{ij}B_{r+s}
\label{N=2}
\\      &  [B_k,B_n]_-=0
\nonumber
\\      &  [B_k,L_n]_-=k B_{k+n}
\nonumber
\\      &  [B_k,G^i_r]_-=-\epsilon^{ij}G^j_{k+r}
\nonumber
\end{eqnarray}
and also the Casimir operator of each of
the (super) $sl(2,\R)$ subalgebras:
\begin{eqnarray}
\N=0:   & L^2   & = L_0^2+L_0-L_1 L_{-1}
\nonumber
\\
\N=1:   & _1 C^2 & = L^2-\frac{1}{2}L_0+\frac{1}{2}G_{1/2}G_{-1/2}
\\
\N=2:   & _2 C^2 & = L^2-L_0+\frac{1}{2}G^i_{1/2}G^i_{-1/2}+\frac{1}{4}B^2
\nonumber
\end{eqnarray}
In~\cite{Kumar:1999fx} 
the quantization of the bosonic classical generators was obtained by
a canonical coordinate transformation:
\begin{eqnarray}
q=\frac{p}{2x} &   &  y=x^2 \, ,
\end{eqnarray}
and setting $f(u)=xp$. The same procedure applied to the $\N=2$ 
generators~(\ref{int_gen}) gives
\begin{eqnarray}
\label{kumar}
\nonumber
G_r^i & = & \psi^i y^{\frac{1}{2}+r}
q^{\frac{1}{2}}
\\
L_n   & = & y^{1+n}q
\\
\nonumber
B_k   & = & y^k B \quad . 
\end{eqnarray}
There are two problems we have to address before quantization. The 
first is how to quantize $\sqrt{q}$ and the second is how to order 
the operators. For the square root of the momentum we define 
the covariant derivative 
$\D^i=\frac{\partial}{\partial \theta^i}+\theta^i \,q$ which 
obey $\{D^i,D^j\}=2\delta^{ij}q$ and since the generators are linear 
in momentum, ordering $y^{n+1}q$ might add a term proportional to $y^n$. 
This motivate the following quantum ansatz
\begin{eqnarray}
\label{quantum_rep}
\nonumber
\sqrt{i}G_r^i & = &  y^{r+\frac{1}{2}}\D^i-i
(r+\frac{1}{2})y^{r-\frac{1}{2}}\theta^iT 
\\
i L_n   & = & y^{1+n}q -i\frac{(n+1)}{2}y^n T
\\
\nonumber
i B_k   & = & y^k B \, ,
\end{eqnarray}
which satisfied the $\N=2$ 
super Virasoro algebra for any $n,k\in \Z$ and $r\in \Z+1/2$ 
provided
\begin{eqnarray}
T & = & \theta^i\frac{\partial} {\partial \theta^i}
\\ 
B & = & \theta^1\partial_2-\theta^2\partial_1
\end{eqnarray}
and which can be truncated to the $\N=1$ algebra generators:
\begin{eqnarray}
\label{quantum_rep_n1}
\nonumber
\sqrt{i}G_r & = &  y^{r+\frac{1}{2}}\D
\\
i L_n   & = & y^{1+n}q -i\frac{(n+1)}{2}y^n T \, ,
\end{eqnarray}
where $T=\theta\frac{\partial}{\partial \theta}$ 
and $\D=\frac{\partial}{\partial \theta}+\theta \,q$. 

In~\cite{Kumar:2000at} a general construction of two half-Virasoro
algebras was given provided $L_1$ and $L_{-1}$ are invertible. The 
conditions for combining them into a single Virasoro algebra were
\begin{eqnarray}
L_1=L_0 L_{-1}^{-1}L_0 &     &  L_{-1}=L_0 L_1^{-1}L_0 \, .
\end{eqnarray}
These conditions are equivalent, when $L_{\pm 1}$ are invertible, to the 
vanishing of the quadratic Casimir
of the $sl(2,\R)$ algebra
\begin{eqnarray}
L^2=L_0^2+L_0-L_1 L_{-1}=0 \, .
\end{eqnarray}
Moreover if we consider the whole tower of $sl(2,\R)$ subalgebras contained
in the Virasoro algebra, i.e 
$\{\frac{1}{n}L_n,\frac{1}{n}L_0,\frac{1}{n}L_{-n}\}$ for $n>0$ with 
the solution found in~\cite{Kumar:2000at}
\begin{eqnarray}
\label{kumar_n=0}
L_n=(L_0 L_{-1}^{-1})^n L_0
\\
L_{-n}=(L_0 L_1^{-1})^n L_0
\end{eqnarray}
we can easily verify that all the quadratic Casimir vanish
\begin{eqnarray}
L^2(n)=\frac{1}{n^2}L_0^2+\frac{1}{n}L_0-\frac{1}{n^2}L_n L_{-n}=0
\end{eqnarray}
We would like to find a similar construction for a representation 
of the super Virasoro generators which is independent of the specific 
super conformal system for the $\N =1$ case~(\ref{N=1}). For this 
we first note that the quadratic Casimir can be written as
\begin{eqnarray}
\label{c2}
_1 C^2 & = & L_0^2+\frac{1}{2}L_0-L_1 L_{-1}+\frac{1}{2}G_{1/2} G_{-1/2}
\\
\nonumber
    & \equiv & \left[G_{1/2} G_{-1/2}-(L_0 +\frac{1}{4})\right]^2-\frac{1}{16}
\end{eqnarray}
and that the realization of the generators in ~(\ref{quantum_rep_n1})
can be thought of as a change of variables which we can invert to obtain
\begin{eqnarray}
q & = & i L_{-1}
\nonumber
\\
\nonumber
T & = & 2(G_{1/2}G_{-1/2}-L_0)
\\
\nonumber
y & = & G_{1/2}G_{-1/2}L_{-1}^{-1}
\\
D & = & \sqrt{i}G_{-1/2}
\\
\nonumber
\theta & = & \frac{2}{\sqrt{i}}(G_{1/2}G_{-1/2}-L_0)G^{-1}_{-1/2}
\\
\nonumber
\frac{\partial}{\partial\theta} & = & -2\sqrt{i}G_{-1/2}(G_{1/2}G_{-1/2}-L_0) 
\, ,
\end{eqnarray}
with $_1C^2=0 \, \Longleftrightarrow \,\theta^2=0$.
This inversion allows us to rewrite the generators of half the $\N=1$ Virasoro 
algebra as functions of only the (super) $sl(2,\R)$ subalgebra generators:
\begin{eqnarray}
L_n & = & \left(G_{1/2}G_{-1/2}L_{-1}^{-1}\right)^{n+1}L_{-1}-(n+1)
\left(G_{1/2}G_{-1/2}L_{-1}^{-1}\right)^n(G_{1/2}G_{-1/2}-L_0)
\nonumber
\\
G_r & = & \left(G_{1/2}G_{-1/2}L_{-1}^{-1}\right)^{r+1/2}G_{-1/2}
\end{eqnarray}
where $n\ge -1$. 
If $L_1$ is also invertible then this solution is valid 
for $n<-1$ and obeys the full algebra. 
The solution reduces to~(\ref{kumar_n=0}) for 
$G_{1/2}G_{-1/2}\sim L_0$. In a similar way to the $\N=0$ 
case, we can compute the quadratic Casimir for all the 
(super) $sl(2,\R)$ subalgebras generators
\begin{eqnarray}
\frac{1}{2k+1}L_{2k+1} \quad,\quad  \frac{1}{2k+1}L_0 \quad, \quad 
\frac{1}{2k+1}L_{-2k-1}
\\
\nonumber
\frac{1}{\sqrt{2k+1}}G_{k+1/2}\quad,\quad  \frac{1}{\sqrt{2k+1}}G_{-k-1/2}
\end{eqnarray}
The quadratic Casimir of the subalgebras are fix, i.e $L^2(k)=L^2$ 
and $_1 C_2(k)=\, _1C_2=0$.
\newpage
\section{Conclusions}
We construct the $\N=1,2$ super Virasoro algebra out of the 
superconformal generators at the classical level of Poisson brackets. 
The generators are ordered and quantized. These quantum generators 
define new coordinates that are inverted and used to construct half 
of the super $\N=1$ Virasoro algebra provided that the quadratic 
Casimir vanish, and the Hamiltonian $H=L_{-1}$ is invertible. This 
condition amounts to the requirement that $H$ has no ground 
state at $E=0$ as is the case in~\cite{Fubini:1984}.
Equivalently, one can demand that supersymmetry is broken 
and there are no states which are annihilated by the 
supersymmetric generators. The full super Virasoro algebra is obtained 
when the special conformal generator $K=L_1$ is also invertible. 

It would be interesting to find out what are the restrictions that higher 
supersymmetry~\cite{Gates:1998ss} put on such constructions and if there
are modifications that will account for central charges, since 
in the $\N=1$ case the representation is not restricted only 
to quantum mechanics in which one would expect these extensions to vanish.

\section{Acknowledgement}
I gratefully acknowledge J. Sonnenschein, M. Kroyter U. Fuchs, and
N. Kashman for useful discussions. This work was supported in part 
by the US-Israel Binational Science Foundation, by GIF - the
German-Israeli Foundation for Scientic Research, and by the Israel 
Science Foundation.

\bibliographystyle{JHEP}
\providecommand{\href}[2]{#2}\begingroup\raggedright\endgroup

\end{document}